\documentclass{revtex4}
\usepackage{amsfonts}
\usepackage{amsmath}
\usepackage{amssymb}
\usepackage{multirow}
\usepackage{bm,graphicx}
\usepackage{wrapfig}

\begin{document}

\title{Nonsymmetrized Hyperspherical Harmonics
approach to $A=6$ system.
\footnote{Presented at the 21st European Conference on Few-Body
Problems in Physics, Salamanca, Spain, 30 August - 3 September 2010.}}
\author{M. Gattobigio}
\affiliation{Universit\'e de Nice-Sophia Antipolis, Institut Non-Lin\'eaire de
Nice,  CNRS, 
1361 route des Lucioles, 06560 Valbonne, France }
\author{A. Kievsky}
\author{M. Viviani}
\affiliation{Istituto Nazionale di Fisica Nucleare, Largo Pontecorvo 3, 56100 Pisa, Italy}

\begin{abstract}
The Hyperspherical Harmonics basis, without a previous
symmetrization step, is used to calculate binding energies of the 
nuclear $A=6$
systems using a version of the Volkov potential acting only on $s$-wave. The
aim of this work is to illustrate the use of the nonsymmetrized basis to deal
with permutational-symmetry-breaking term in the Hamiltonian, in the present
case the Coulomb interaction.
\end{abstract}

\maketitle
\section{Introduction}
\label{intro}
The Hyperspherical Harmonics (HH) basis set has been extensively used 
to describe bound states and scattering processes
in $A=3,4$ systems~\cite{report}. One of the main reasons of using the
HH basis resides in its flexibility, however it suffers from a large
degeneracy, which, up to now, has prevented a systematic use with realistic
potentials for $A\ge 4$ systems. 
The authors have recently proposed an approach \cite{gatto1,gatto2,gatto3} in which the
HH basis is not symmetrized. In particular, in Ref.~\cite{gatto3} the nonsymmetrized
basis has been used to describe systems up to $A=6$ particles. It was shown that the 
large degeneracy of HH can be tackled noticing that
the Hamiltonian can be expressed as an algebraic
combination of sparse matrices, and the diagonalization procedure was implemented by
means of an iterative diagonalization, where only the action of the Hamiltonian on
a vector is required. 
In this work we continue the study on the $A=6$ system, interacting {\it via} a
two-body Volkov potential acting in $s$-wave,
analyzing how the introduction of Coulomb interaction breaks the original 
permutation symmetry $S_6$, in the case of two or three protons.

\section{The method} 
\label{sec:1} 
Following Ref.~\cite{gatto1}, we introduce the Jacobi
coordinates, $\mathbf x_1,\dots,\mathbf x_N$, with $N=A-1$,
and we ``adapt'' the coordinates to the particle pair $(i,j)$
defining $\mathbf x_N=\mathbf r_i-\mathbf r_j$, where $\mathbf r_i$ and $\mathbf
r_j$ are the Cartesian coordinates of particles $i$ and $j$.
From the Jacobi coordinates, we introduce  the hyperspherical coordinates, that
means an hyper-radius $\rho$ and $3N-1$ hyperangles $\Omega_N^{ij} = (\hat x_1,
\dots, \hat x_N, \phi_2, \dots, \phi_N) $.

The HH functions with fixed angular momentum $LM$, ${\mathcal
Y}^{LM}_{[K]}(\Omega_N^{ij})$, are defined as the eigenfunctions of the grand angular 
momentum. They are labelled by the grand angular quantum number $K$
and a set of $3N-1$ quantum numbers indicated by $[K]$. In Ref.~\cite{gatto3}
it was shown that the potential energy, at fixed values of the hyper-radius, can be written as
\begin{equation}
\sum_{ij} V_{ij}(\rho)=\sum_{ij} 
[{\cal B}^{LM}_{ij}]^t\, V_{12}(\rho)\,{\cal B}^{LM}_{ij} \,,
\label{eq:vpot}
\end{equation}
where the potential matrix 
$[V_{12}(\rho)]_{[K'][K]}= \langle{\cal Y}^{LM}_{[K']}(\Omega^{12}_N)|V(1,2)|{\cal
Y}^{LM}_{[K]}(\Omega^{12}_N)\rangle$ is a sparse matrix and the matrix 
${\cal B}^{LM}_{ij}$ is given by a product of sparse matrices
corresponding to the transformation of the HH vector defined on $\Omega_N^{ij}$ to
the one defined on
$\Omega_N^{12}$. The potential energy matrix is obtained after integrating on $\rho$ using
a Laguerre basis. Furthermore, the kinetic energy is diagonal in HH basis, and
this allows to write the full Hamiltonian as an algebraic combination of sparse
matrices. The diagonalization of the Hamiltonian is obtained {\it via} an
iterative scheme, where only the action of the Hamiltonian on a vector is
required.

\section{Results}
For our numerical application, we chose a Volkov potential
\begin{equation}
 V(r)=V_R \,{\rm e}^{-r^2/R^2_1} + V_A\, {\rm e}^{-r^2/R^2_2} \,,
\end{equation}
with $V_R=144.86$ MeV, $R_1=0.82$ fm, $V_A=-83.34$ MeV, and $R_2=1.6$ fm, which
only acts in $s$-wave, and with the mass such that $\hbar^2/m =
41.47~\text{MeV\,fm}^{2}$. We have calculated the binding energy for a system
of $A=6$ particles, and we repeated our calculations for the same system with
Coulomb interaction between two particles, {\it i.e.} a model of
$\,^6\text{He}$, and between three particles, {\it i.e.} $\,^6\text{Li}$.

In Table~\ref{tab} we show the results for the $L=0$ state. Without the
Coulomb interaction the symmetry
group is $S_6$ and to antisymmetrize the wave function, taking also
into account the spin and isospin degree of freedoms, the eigenvalue must belong
to the irreducible representation $[\mathbf 4\,\mathbf 2]$. We want to stress
the fact that, using the nonsymmetrized basis, 
the eigenvectors belong to all of the irreducible representations of $S_6$, and
in particular that of interest. 

When we add Coulomb interaction between two particles, the symmetry is broken as
$S_6 \rightarrow S_2\otimes S_4$, and the original level, having degeneracy
9, is split into 4 sub-levels 
\begin{equation}
  [\mathbf 4\, \mathbf 2]  \rightarrow [\mathbf 1^2]\otimes[\mathbf 3\, \mathbf
  1] + [\mathbf 2]\otimes[\mathbf 4] + 
  [\mathbf 2]\otimes [\mathbf 3\,\mathbf 1] + [\mathbf 2]\otimes[\mathbf 2^2] \,,
  \label{}
\end{equation}
where only $[\mathbf 2]\otimes[\mathbf 2^2]$ is physical, as is the only one
that can be antisymmetrized with respect the four neutrons using the spin degree
of freedom, and describes the ground state of $\,^6\text{He}$. When the Coulomb
interaction is extended to three particles, the symmetry breaking is
$S_6 \rightarrow S_3\otimes S_3$, and the split reads
\begin{equation}
  [\mathbf 4\, \mathbf 2]  \rightarrow [\mathbf 2\,\mathbf 1]\otimes[\mathbf 3] +
  [\mathbf 2\,\mathbf 1]\otimes[\mathbf 2\,\mathbf 1] + 
  [\mathbf 3]\otimes [\mathbf 3] + [\mathbf 3]\otimes[\mathbf 2\,\mathbf 1] \,,
  \label{}
\end{equation}
with the only physical state, describing $\,^6\text{Li}$, being $[\mathbf
2\,\mathbf 1]\otimes[\mathbf 2\,\mathbf 1]$.

We have shown the power and the flexibility of the
nonsymmetrized HH approach. In particular we were able to include basis states up
to $K=22$ for a six-body system (corresponding to a basis set of
$38^.798^.760$ elements using 11 Laguerre polynomials ). Furthermore,
using a symmetry-adapted Lanczos algorithm, we were able to trace the
irreducible representation of the eigenvector of interest and to select the
corresponding eigenvalue. 
\begin{table}
  \caption{Binding energies calculated with Volkov's potential in $s$-wave, for
  $A=6$ particles and $L=0$, with and without Coulomb interation, using 11
  Laguerre's polynomials.  }
  \label{tab}
\begin{center}
  \begin{tabular*}{0.9\linewidth}{@{\extracolsep{\fill}}c c c c c}
  \hline 
  \hline 
  $K_{\text{max}}$ \rule{0pt}{12pt} &  $N_{\text{HH}}$   & $E_0$ (MeV)  &
  $E_{[\,^6\text{He}]}$ (MeV) & $E_{[\,^6\text{Li}]}$ (MeV) \\
  & & [\bf{4}\,\bf{2}] & $[\mathbf{2}]\otimes[\mathbf{2}^2]$ &
  $[\mathbf{2\,1}] \otimes [\mathbf{2\,1}]$   \\
  \hline \\
   2 & 15   &    24.793 & 24.064 & 22.974 \\
   4 & 120  &    28.791 & 28.016 & 26.988 \\
   6 & 680  &    30.723 & 29.935 & 28.947 \\
   8 & 3045 &    31.645 & 30.851 & 29.889 \\
   10 & 11427 &  32.244 & 31.446 & 30.496 \\
   12 & 37310 &  32.708 & 31.908 & 30.964 \\
   14 & 108810 & 33.075 & 32.275 & 31.334 \\
   16 & 288990 & 33.358 & 32.558 & 31.620 \\
   18 & 709410 & 33.561 & 32.762 & 31.827 \\
   20 & 1628328 &33.710 & 32.912 & 31.980 \\
   22 & 3527160 &33.814 & 33.016 & 32.087 \\
   \hline
\end{tabular*}
\end{center}
\end{table}


\end{document}